# THE STRUCTURE OF A MELT: THE CASE OF LIQUID BISMUTH

Flor B. Quiroga [1], Isaías Rodríguez [1], David Hinojosa [2], Alexander Valladares [2], Renela M. Valladares [2] and Ariel A. Valladares [1]*

[1] Instituto de Investigaciones en Materiales, Universidad Nacional Autónoma de México, Apartado Postal 70-360, Ciudad Universitaria, CDMX, 04510, México.

[2] Facultad de Ciencias, Universidad Nacional Autónoma de México, Apartado Postal 70-542, Ciudad Universitaria, CDMX, 04510, México.

* Corresponding Author, Ariel A. Valladares, valladar@unam.mx.

**Abstract**

Molecular Dynamics (MD) is performed on supercells of 216 atoms of bismuth, going from 300 K to 573 K in 100 steps and maintaining it in the liquid state, at 573 K, during 500 steps using the Materials Studio (MS) suite of codes. The Pair Distribution Functions (PDFs) and the Plane Angle Distributions (PADs) of the last 1, 10, 25 and 100 steps of the MD have been obtained. Averaging the last 100 steps, as representative of the liquid, Reverse Monte Carlo (RMC) was applied to obtain 4 atomic structures, one for each set of initial random velocities. Then, a detailed structural study of liquid bismuth at 573 K was undertaken; PDFs and PADs are calculated and reported. Two noticeable peaks appear for the PDFs, at 3.25 and 6.55 Å, along with a pseudo peak (shoulder) at 4.6 Å. This shoulder (after the first peak) of the PDFs is found to be related to the third and fourth neighbor peaks of the crystalline Wyckoff structure and also to the diagonal distances in deformed squares in the liquid structure. For completeness J(r)s are also reported. Two prominent peaks in the MS PADs are observed: 53° and 85°; and two for the RMC PADs: 58° and 90°, suggesting the existence of deformed triangles and squares. Less abundant are higher-order geometrical structures.

## 1. INTRODUCTION

Bismuth (Bi), both in its solid and liquid phases, is one of the materials better studied. Its attraction may have started when it was discovered that it has a low melting temperature [1], 544 K, and the appealing geometrical and colorful forms when it crystallizes, the friezes are reminiscent of Mayan ruins. Fig. 1. Whatever the reason, Bi is an interesting element of the periodic table that has important applications in the industrial world, [1-5]. But despite its well-known properties, it still is giving surprises and properties to investigate, due to its versatility. In its crystalline state, at room temperature and atmospheric pressure (the Wyckoff structure), it is a semimetal; however, when it undergoes a transformation from the crystalline to the amorphous state, it changes and becomes metallic, [6-8] . This metallic property is retained when it melts, [6, 8-9]

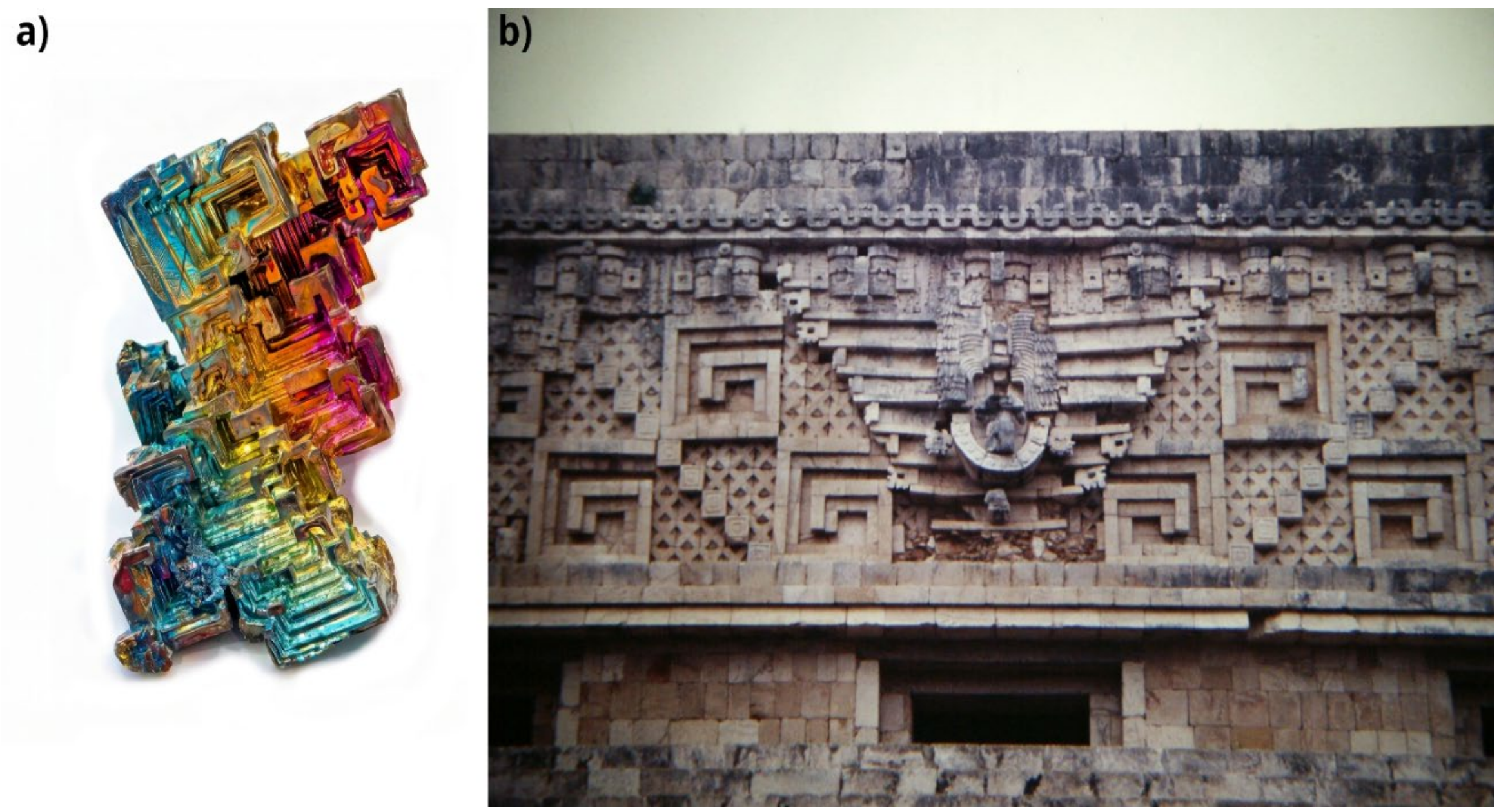

*Figure 1. Comparison of complex crystalline and cultural structures. a) Bi crystal showing the colorful pattern and the geometrical hauts-reliefs of the mineral. b) Detail of the "Casa del Gobernador" façade in Uxmal, Yucatán, Mexico showing the hauts-reliefs. (Original image by Gary Todd, via Wikimedia Commons).*

Solid bulk phases in Bi extend to high pressures with the corresponding changes in the crystalline structures, Fig. 2. At the beginning, there were several structures that were not stable but nevertheless were considered as true phases [10]; further experimental work isolated with better precision which ones were real phases and presently, [10] five phases (Bi I, the Wyckoff structure; Bi II, monoclinic structure; Bi III, tetragonal structure proposed by Chen *et al.* [11] ; Bi IV, orthorhombic; and Bi V, body-centered cubic) have been identified with well-defined structures, (except for the Bi III phase that is considered incommensurable). These structures appear for pressures lower than 7 GPa and, incidentally, they are superconductors, experimentally determined, (except for the Bi IV phase that was predicted by our group to superconduct at a transition temperature of 4.5 K [12]). But they are not the only ones that superconduct; also the amorphous phase is a superconductor with a transition temperature ten thousand times larger (about 6 K, [13-14]) than the Wyckoff structure (0.53 mK, [15]), Also, it has been observed that whenever Bi participates in alloys or compounds, their tendency is to become superconductive, [16-26].

All these results seem to indicate that structural defects, like disorder, increase the propensity to superconduct. In a previous publication [27] we analyzed some of the atomic arrangements that appear in amorphous Bi to try to investigate if these structures may account for this superconductive behavior. So the question remains unanswered: are there specific atomic arrangements that may foster heightened electronic properties and high Tc superconductivity? These specific arrangements may appear in the liquid.

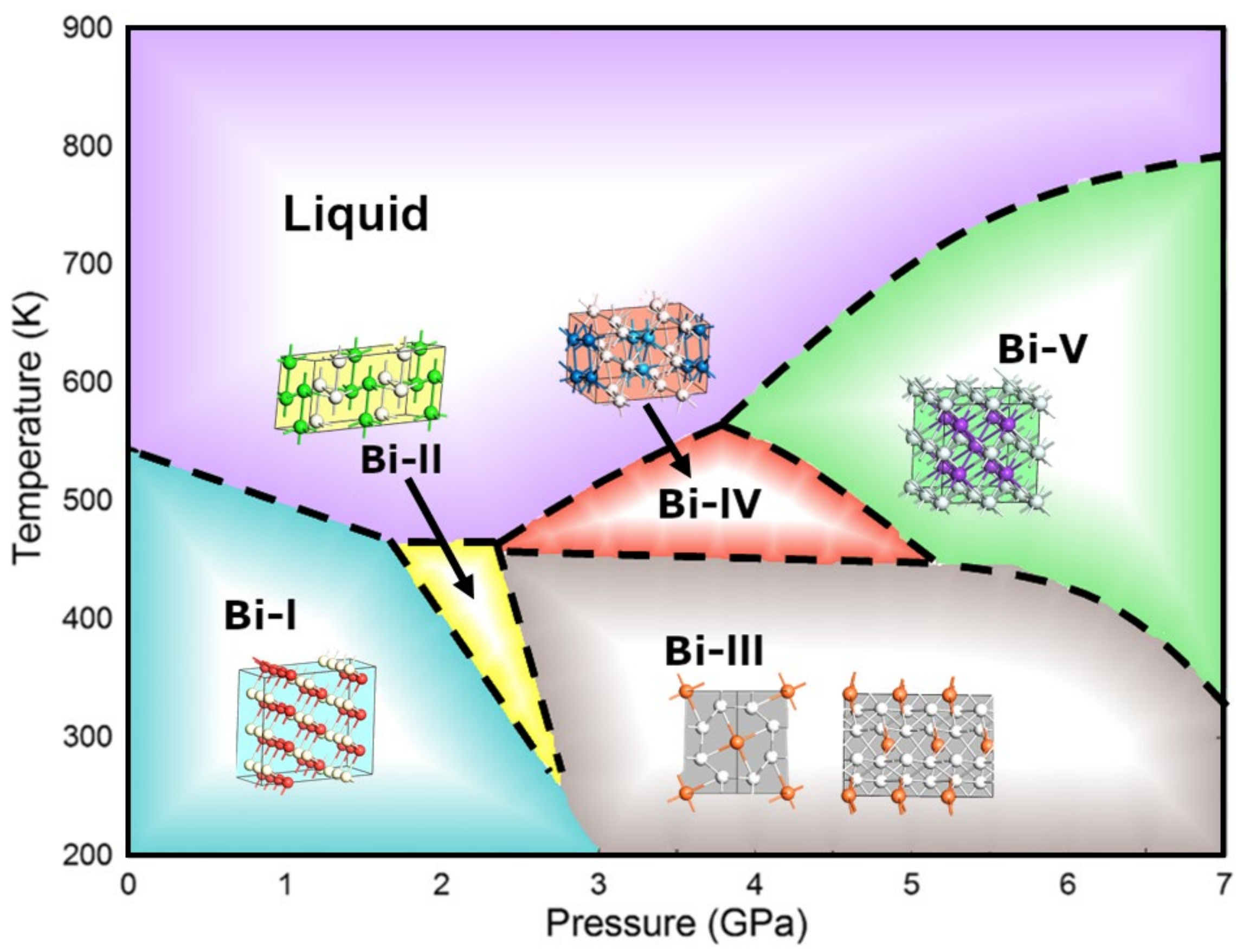

*Figure 2. Phases of solid bismuth with the corresponding crystal structures. The insets are the crystalline structures observed. For a description of the phases see text. This diagram is a modification of the one in Ref [28].*

Its liquid phase has been the subject of topological studies to identify possible arrangements that hint at the existence of peculiar local structures. There are discrepancies, both simulational and experimental, concerning the shapes of both the PDFs and the PADs. Some authors argue that the existence of a conspicuous pseudo peak (shoulder) in the PDFs of the liquid is due to specific atomic structures [6, 29] while others consider that the experimental procedure of measurement is responsible for the appearance of this feature, [30-32]. Concerning the PADs, discrepancies exist in the literature between the positions of the prominent peaks reported. Recent work ([29] and references contained therein), reports the existence of specific structures that give rise to the observed shoulder in the PDFs and relates them to certain prominent angles in the PADs. In particular, they find that the prevalent structures are triads of atoms bonded into triangular chains with plane angles around 45º and 90º, and when these triads are removed, the shoulder disappears. No definite conclusions yet [29].

So it would be natural to ask if some propitious structures may be found in liquid Bi to enhance its electronic properties even more. But, in the process, it is also important to investigate if we can contribute to elucidate the different viewpoints of the atomic structure of liquid Bi. Stimulated by these questions, a study of atomic arrangements in liquid Bi was undertaken. This is the origin of the present work.

## 2. METHODOLOGY

To study all these features, four superlattices were constructed and computationally liquefied; each of these superlattices contained 216 atoms as a result of multiplying the unit cell of a diamond-like structure (3 x 3 x 3) times. The diamond structure was chosen as an initially unstable structure of Bi to propitiate the evolution into a disordered topology, maintaining the experimental density of 10.029 g/cm$^3$ for the liquid structure at 573 K. To see if the structures generated were dependent on the initial random distribution of velocities, each of four samples was started with a different velocity set, identified as $v_0$, $v_1$, $v_2$, $v_3$, with a lattice parameter of 19.5516 Å, which gives the experimental density mentioned above.

To generate the liquids, *ab initio* Molecular Dynamics (MD) was used based on a Density Functional approach (DFT) implemented in the DMol$^3$ code [33], part of the Materials Studio Suite [34]. The MD process started at 300 K linearly heating the samples to 573 K (29 K above the melting temperature of Bi) in 100 steps of 19.4 fs each. After that, the temperature is maintained constant at 573 K for 500 steps. The MD procedure uses a Nosé-Hoover NVT thermostat with a Nosé Q-ratio of 0.5 [35]. The parameters of the electronic calculations were based on a double numerical basis set with d-function polarization (dnd) and a 6.0 Å real-space cutoff. The LDA-VWN functional [36] was used for the computation of the exchange-correlation energy and the Semi-Core Pseudo Potentials (dspp) for the core treatment. All calculations were spin-unrestricted with a Self-Consistent Field (SCF) convergence of 1 x 10$^{-6}$ Ha, using a thermal smearing of 5 x 10$^{-3}$ Ha.

Different initial atomic velocities for each of the four samples were used, which were obtained with ***Velocities***, a code developed by our group [37]. To ensure that the final structures of our supercells correspond to a liquid, the PDFs are obtained and compared with experimental results at different intervals. Using Reverse Monte Carlo (RMC) [38-39], four atomic structures were generated starting from the average PDFs of the last 100 steps in the MD process for each of the four initial velocities, making them more representative of the liquid phase of Bi. To see the consistency of our results, we averaged the last 100 steps and compared them with each other and with results existing in the literature.

Also, the PADs were calculated using ***Correlation***, a code developed in our group, [40] for each of the 4 structures obtained with RMC and for the structures calculated with MS (in this case, for the average of the last 100 steps). It is evident that the angles calculated for the PADs depend on the definition of the bond length, and for this purpose, a pair of atoms is considered bonded when their interatomic distance is, at most, 4.185 Å, the position where the minimum between the first peak and the shoulder is located.

To show the energy variation during the last 500 steps, Fig. 3 represents the values of the energy in this interval and the corresponding slope, intersection and uncertainty to demonstrate that the variation is negligible. This graph is for the velocity $v_1$, but the results are very similar for the other three velocities.

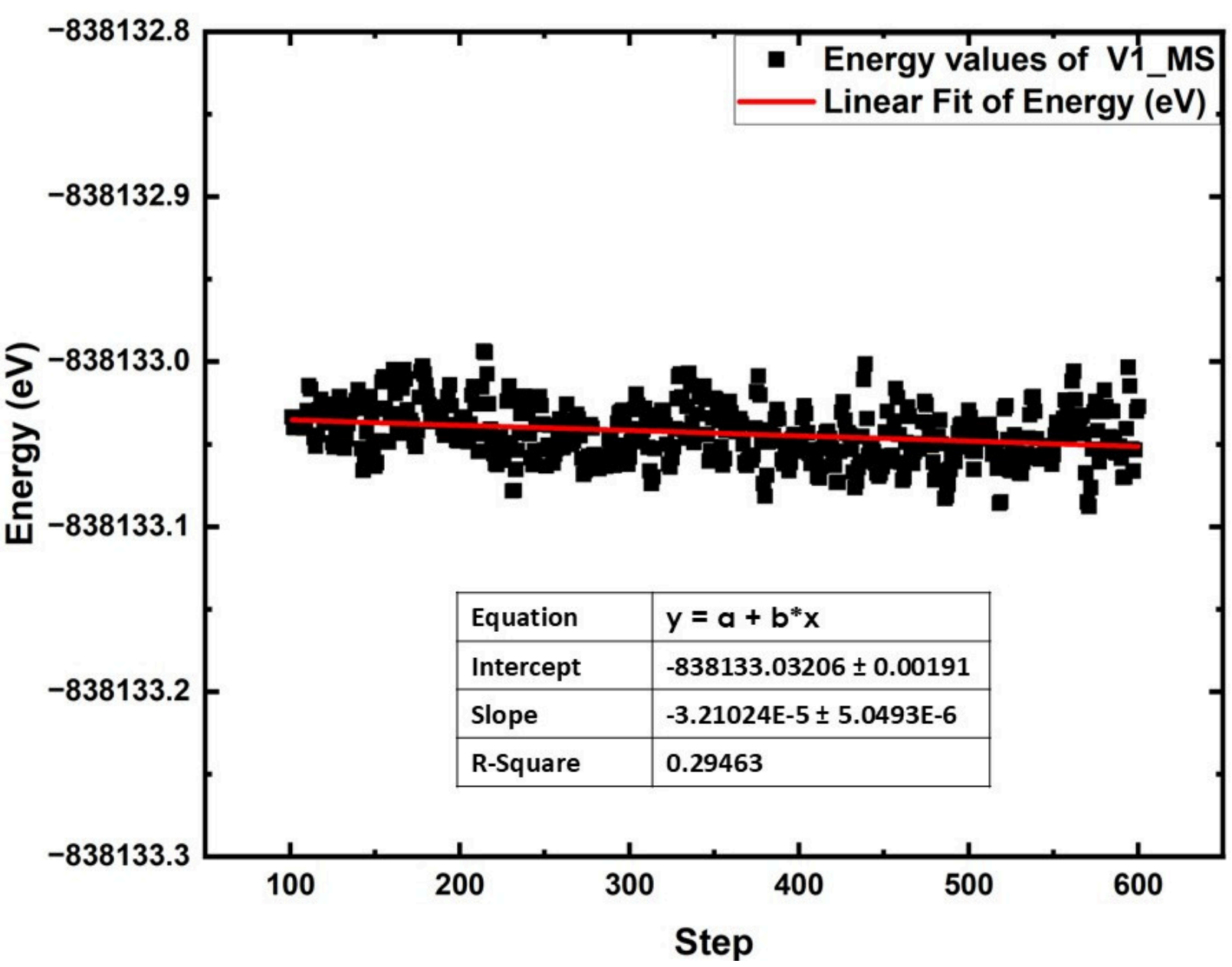

*Figure 3. Energy variation in electron-volts during the last 500 steps of the simulation. It can be seen that the slope of the line is -3.2 x $10^{-5}$, eV/step, with an uncertainty of 5 x $10^{-6}$ eV/step. The variation is negligible.*

Finally, a detailed study of the atomic structures in the simulated liquid was conducted.

## 3. RESULTS

So several questions are relevant: Are the PDFs and PADs obtained in this work comparable with the experimental and simulational ones reported in the literature? What are the most common geometrical structures in the liquid, consistent with the PDFs and the PADs calculated? Is the shoulder related to some specific structure? Is this shoulder the result of some experimental approximations when determining the PDFs from the structure factors? This work aims to answer some of these questions.

### *3.1 Atomic Structures and Topologies*

In Figs. 4, the PDFs calculated with Materials Studio (MS) from 4 different initial velocities, averaged over several number of last steps: 1, 10, 25, and 100, are compared, and it is clear that for the 100-step averages, very similar PDFs are obtained regardless of which initial velocities are used, although they were randomly different.

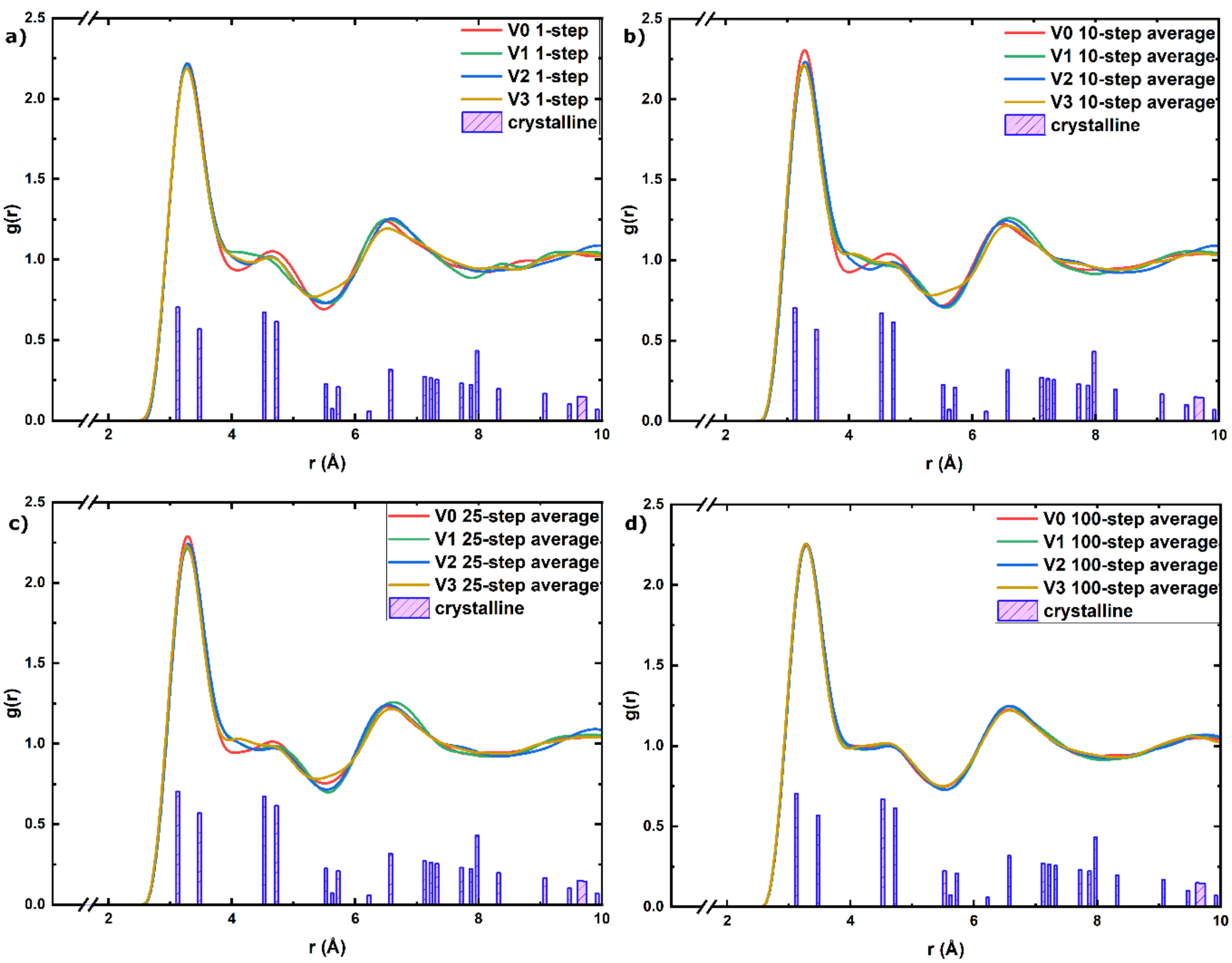

*Figure 4. Comparison of the PDFs generated by MS with four different initial random velocities, showing a) the last step, b) the average of the last 10 steps, c) the average of the last 25 steps, and d) the average of the last 100 steps. In figure 4d) the PDFs are indistinguishable.*

In Table 1 the position of several peaks of the liquid PDFs are registered. Since there are coincidences with the crystalline structure, the positions of the peaks for this structure are also included.

| **Liquids** | **1st peak [Å]** | **Shoulder peak [Å]** | **2nd peak [Å]** |
|---|---|---|---|
| **V0_MS** | 3.25 | 4.55 | 6.55 |
| **V1_MS** | 3.25 | 4.65 | 6.55 |
| **V2_MS** | 3.25 | 4.65 | 6.55 |
| **V3_MS** | 3.25 | 4.55 | 6.55 |
| | **crystal** | | |
| **1st peak [Å]** | **2nd peak [Å]** | **3rd peak [Å]** | **4th peak [Å]** |
| 3.12 | 3.47 | 4.52 | 4.72 |

Table 1. The positions of the first 3 peaks in the PDF's of the four Bi liquids and the first 4 peaks of the crystal are displayed.

Due to the similarities of the 100-step averages, we decided to use these PDFs to generate four structures using RMC to atomically describe the liquid system. It can be observed that there are differences in the PDFs (*g(r)s*) for the 1, 10, and 25 steps cases from 3.5 Å onwards, and despite this, each PDF describes liquid Bi. These differences disappear for the 100-step PDFs. The structures obtained from RMC will be used to calculate the new PDFs and also the PADs. Since our updated ***Correlation*** [40] code can also be used to obtain properties such as PDFs and PADs, we will calculate the aforementioned averages for our MS supercells and also use them in conjunction with our RMC-generated samples. All of our graphs used 5-point FFT smoothing.

The graphs will now change their nomenclature from $V_X$ {y}-step average to $V_X$_MS, which corresponds to the data obtained from MS, and the ones obtained using RMC will be denoted as $V_X$_RMC.

In Figs. 5 the average PDFs of the 100 MS steps are compared with the PDFs of the new supercells obtained from RMC. It can be seen at a glance that the PDFs obtained are identical, so we conclude that by using RMC we are obtaining supercell structures representative of liquid Bi. Using RMC was useful to find the atomic structure of the liquids and in this manner we were able to obtain PDFs and PADs from the start, without the need to average curves. However, for a test of our results, we compare the statistical distribution functions obtained from the RMC structures with the MS average curves obtained from the last 100 steps later. Then we are free to use either the RMC structures or the MS averages shown in Figs. 5.

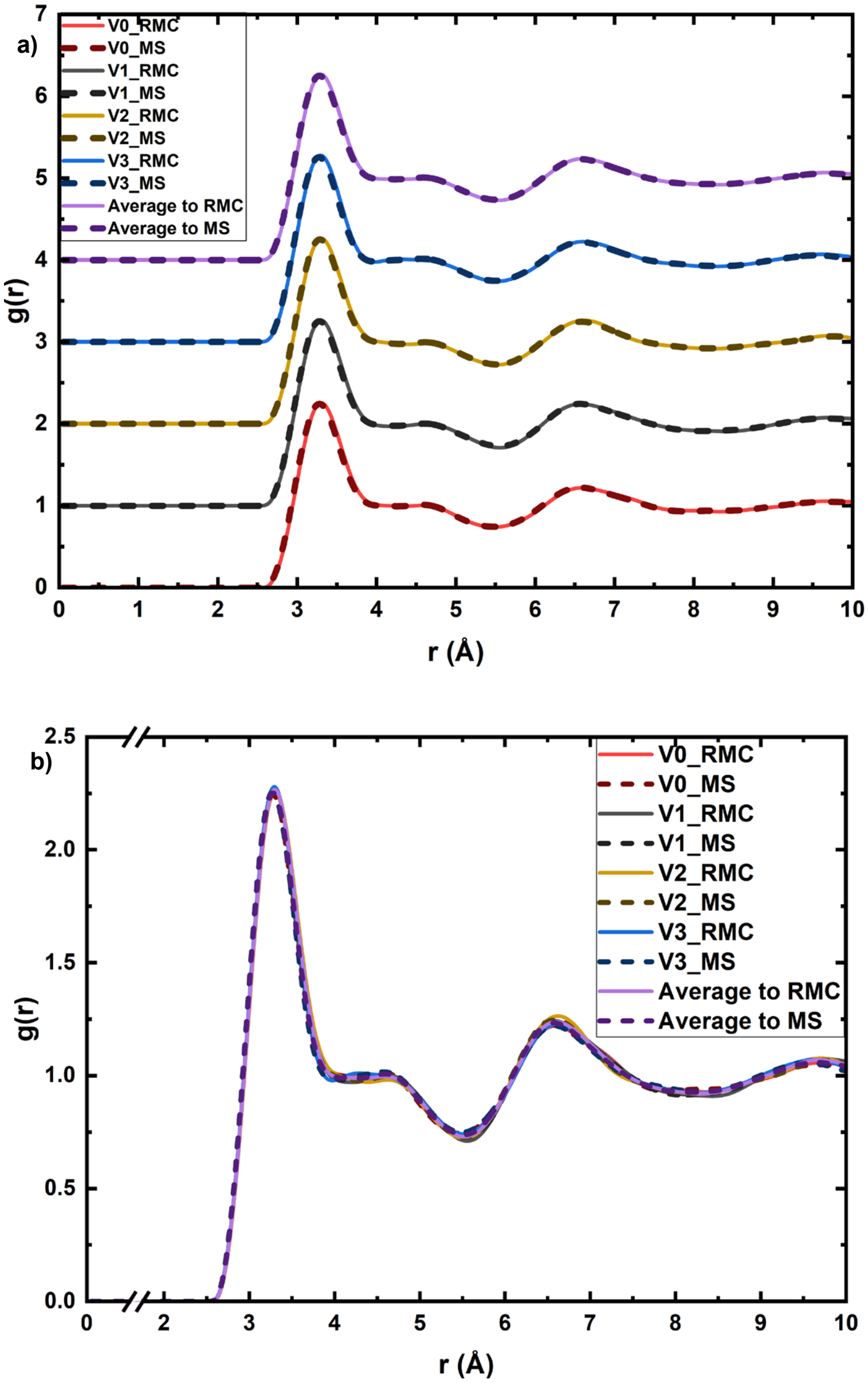

*Figure 5 Comparison between the averages of the PDFs obtained from MS for 4 different velocities (dashed curves) and those from RMC. a) The averages of RMC and MS are also included. The curves are displaced vertically for a better appreciation; b) Curves superimposed; the similarity is evident.*

In Figs. 6 our averaged PDFs of MS and RMC are compared with experimental and computational results. In Fig. 6a) the comparison is with experiments at 553 K. Discrepancies are observed in the experimental peaks, and most notably, in the positions, heights and breadths among the first peaks and among the second peaks. Here, the conspicuous shoulder that characterizes liquid Bi is present, but it should be noted that not all curves display it. In Fig. 6b) our results are compared with experiments at 573 K, where there is a small protrusion where the shoulder should be [29], whereas in Mudry's results [7] there is none. Finally, in Fig. 6c) we compare with other computational results, where the curves match reasonably well in the shoulder region. We can infer that the existence of this shoulder is indeed a structural possibility in the liquid.

When Waseda and Suzuki, [41], in their pioneering work first determined the PDF for liquid Bi they reported the existence of a shoulder in the structure factor,  a shoulder that disappeared when the PDF was calculated. Ulterior experimental studies manifested its existence. When computer simulations started to be relevant for the analysis of these structures the shoulder appeared and stayed.

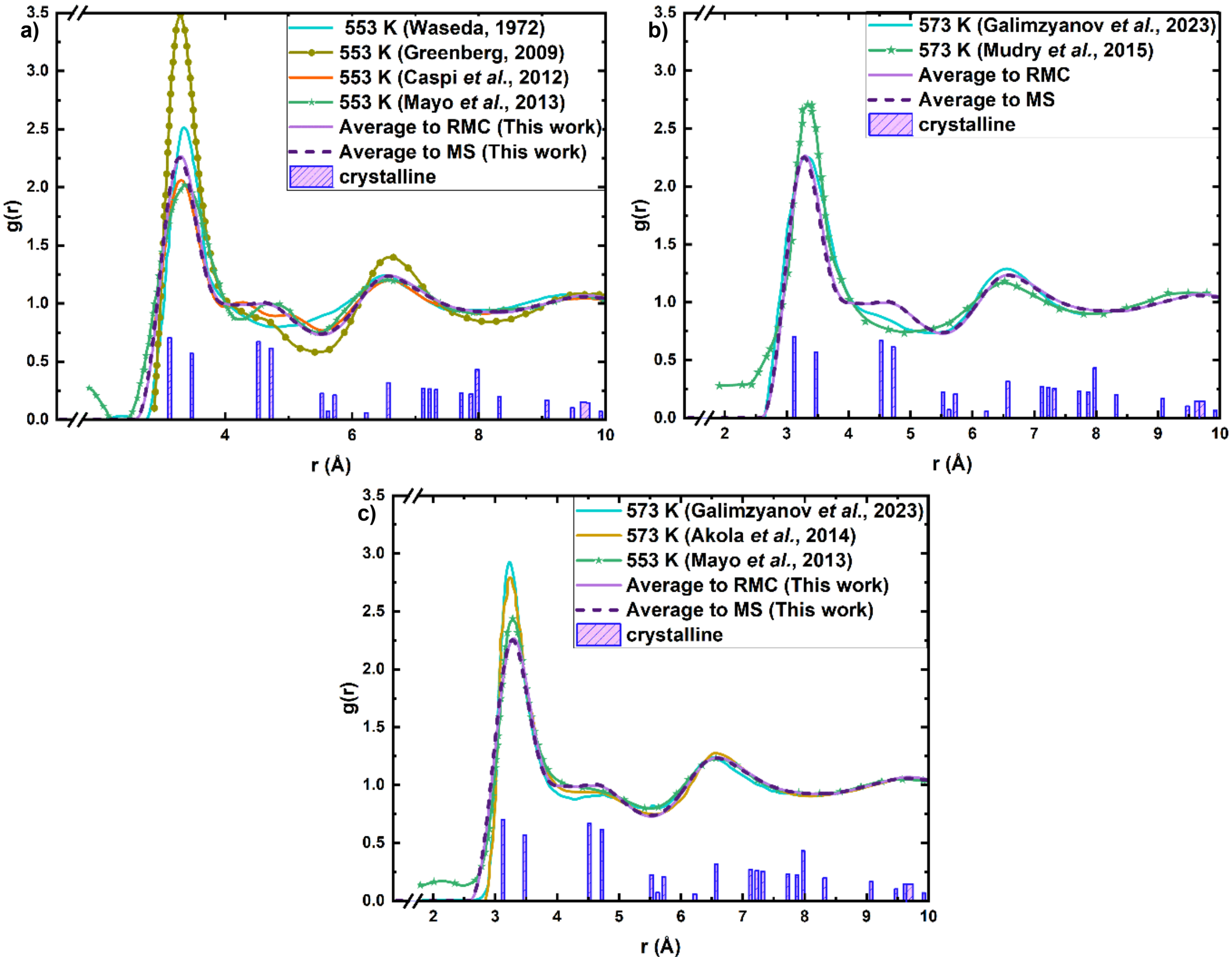

*Figure 6. Comparison of our averaged PDFs obtained with MS and RMC to reported results, both experimental and theoretical. a) Comparison with experimental results (T=553 K) [30, 32, 41-42] b) Comparison with experiment (T=573 K) [7,29] and c) Comparison with other computational works (T=573 K) [8,29-30].*

For completeness, in Fig. 7, the radial distribution functions, J(r)s, of the last 1, 10, 25 and 100 steps are included, for each initial random velocity. The four random velocities lead to practically the same J(r)s, but the last 100 steps reproduce indistinguishable J(r)s. In Fig. 8, we compare our results for RMC and MS and again, the J(r)s are indistinguishable.

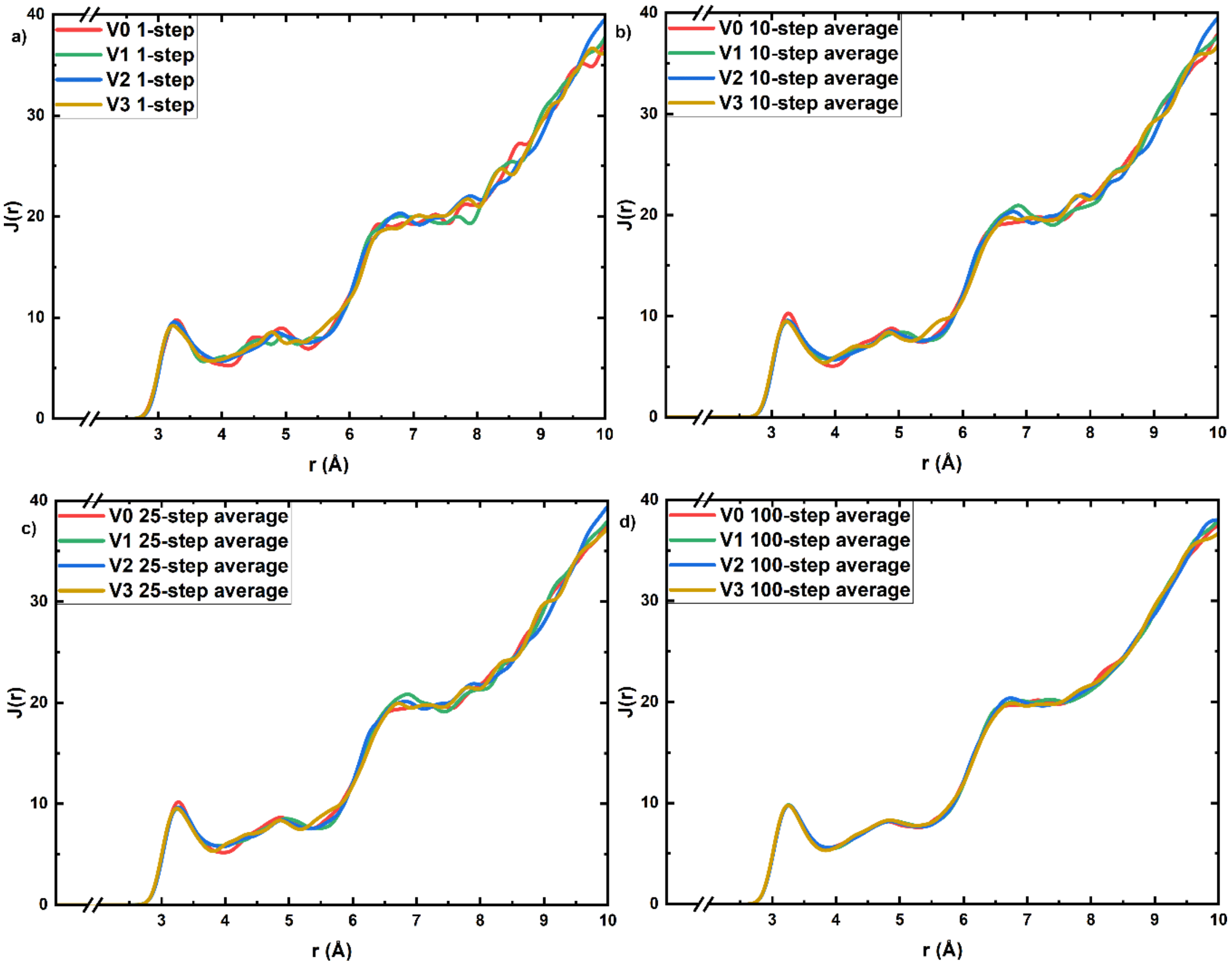

*Figure 7. Radial Distribution Functions for the last 1, 10, 25 and 100 steps for our MS samples. The four initial random velocities lead to practically the same J(r), but the coincidence in 7d) is remarkable.*

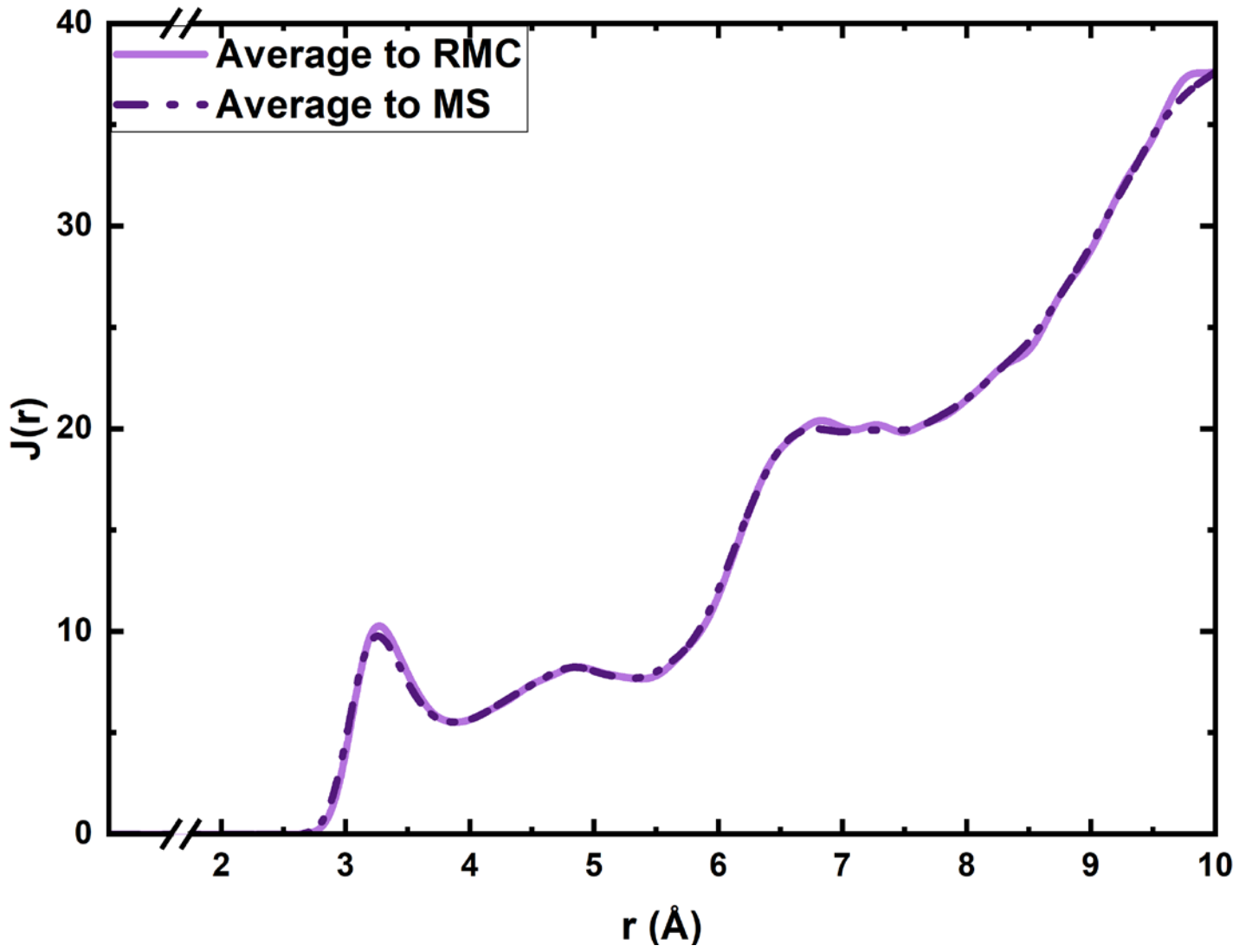

*Figure 8. Comparison of our results obtained with MS and RMC.*

### 3.2 The shoulder; information from PDFs and PADs

From the discussion above, one can see that the first peak in the liquid appears as a consequence of coalescing the first two delta crystalline peaks (3.12 and 3.47 Å), whereas the group of crystalline delta functions located at 4.52 and 4.72 Å resembles the position of the shoulder. It seems that these two peaks coalesce to give rise to the shoulder. However, information from the calculated PADs should also be considered. PADs are less common in the literature since their experimental determination is more difficult. However, there are some simulational calculations that will be compared with our present work. Our calculations of PADs are presented in Figs. 9.

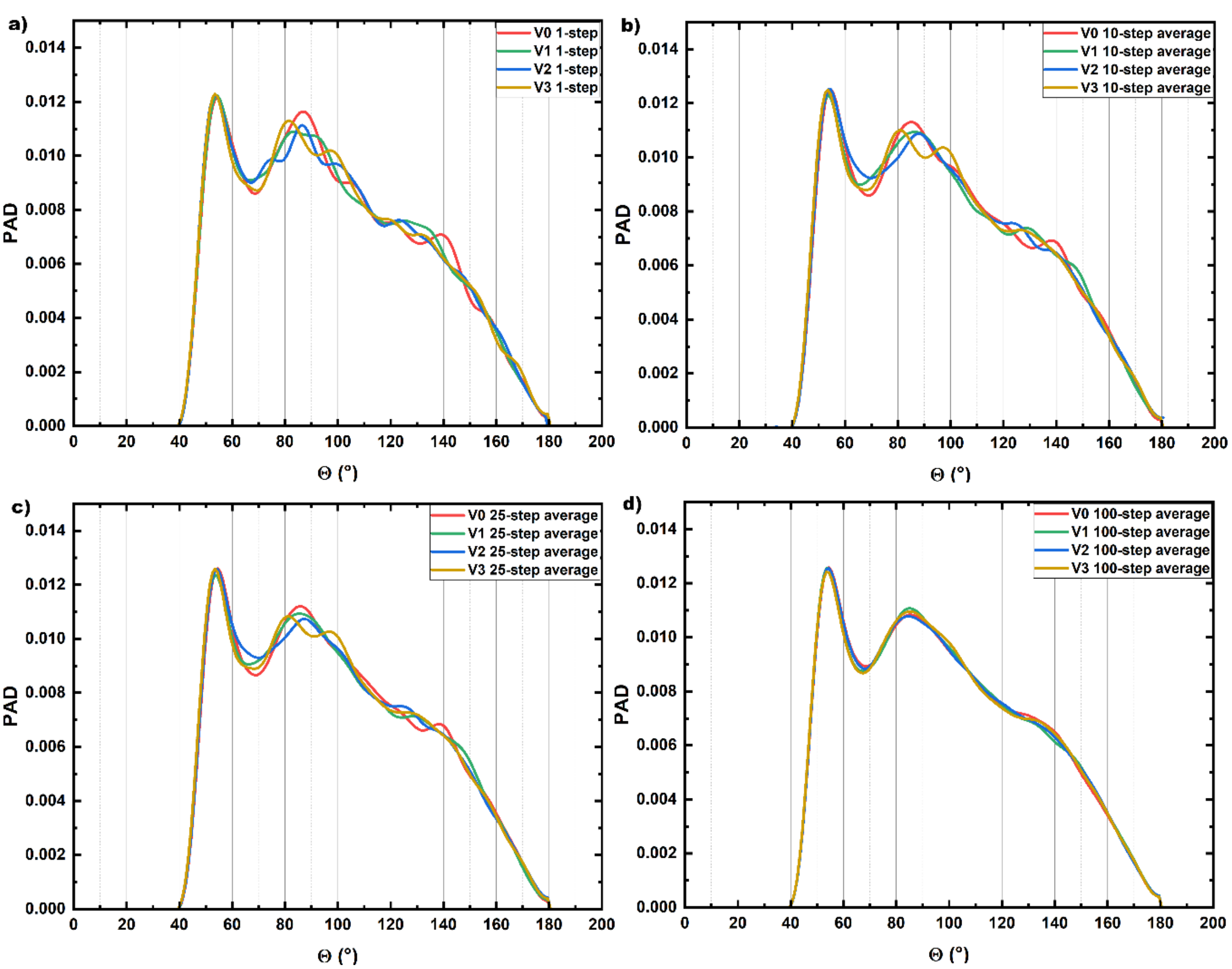

*Figure 9. PADs for the four initial random velocities at 1, 10, 25 and 100 final steps. Angles between 50º and 60º are observed, together with angles around 90º, reminiscent of equilateral triangles and square structures, respectively.*

The position of the shoulder also coincides with the value of the interatomic distance between atoms located at the end of the diagonals of the square structures, revealed by the presence of 90º angles in the calculated PADs, as shown in Figs. 9: ~ 4.6 Å;

see Table 1. Therefore, the shoulder exists because of the coalescence of the two second neighbor crystalline peaks and the existence of square structures.

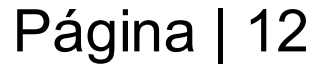

Also, the highly localized counts at about 53° in the PADs of Figs. 9 indicate the presence of atomic triads, suggesting the existence of equilateral triangles (three angles of about 60º). The absence of angles below 45º rules out elongated isosceles triangles in agreement with the information from the PDFs, since the interatomic distance for the shortest side of these triangles does not match the characteristics of the reported PDFs. On the other hand, the appearance of a broad peak in the PADs at about 90º (from 80º to 100º) suggests the existence of distorted atomic squares, or rhombi, with the interior angles varying in this interval. Fig. 10 is a comparison of our averaged results obtained with MS and with RMC and the results reported by Galimzyanov [29] and Akola [8]. In our averages, pronounced peaks appear around 60° and around 90°, and it can be seen that our PADs agree with some previously published.

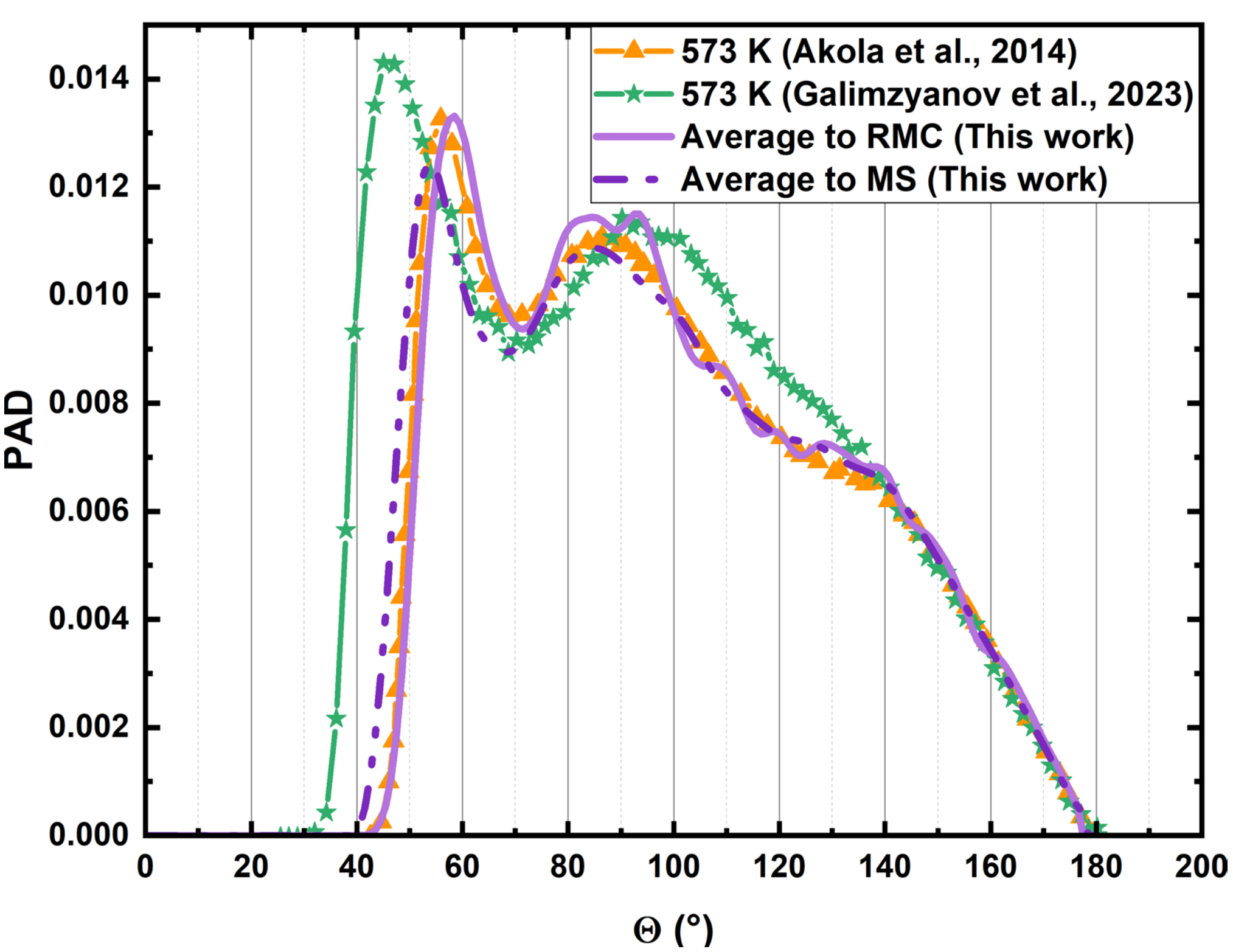

*Figure 10. Comparison of several PADs. Ours correspond to the average of 4 different supercells generated by MS (Average of 100 steps) and by RMC obtained using the **Correlation** code [40], compared with the results of [8, 29].*

The presence of the shoulder has focalized the attention of the community and appears and disappears intermittently. It has been argued [30] that the shoulder appears due to the different precisions with which the structure factor of liquid Bi is determined, and when this factor is Fourier-transformed to obtain the PDFs, the shoulder appears with more or less preponderance in these PDFs. However, the fact that it appears in simulations that do not involve experimental measurements but direct calculations of the PDFs, without dealing with the Structure Factor, seem to indicate that although the structure factor may contribute to the results, it cannot be a determining factor. Based on our detailed analysis of the atomic topology, our simulations reveal the presence of this shoulder at every turn, which indicates that this feature is a property of the atomic distribution in the liquid state.

The presence of other less prominent angles between 120º and 140º should be noted. This indicates the likely existence of more complex atomic structures that should be investigated.

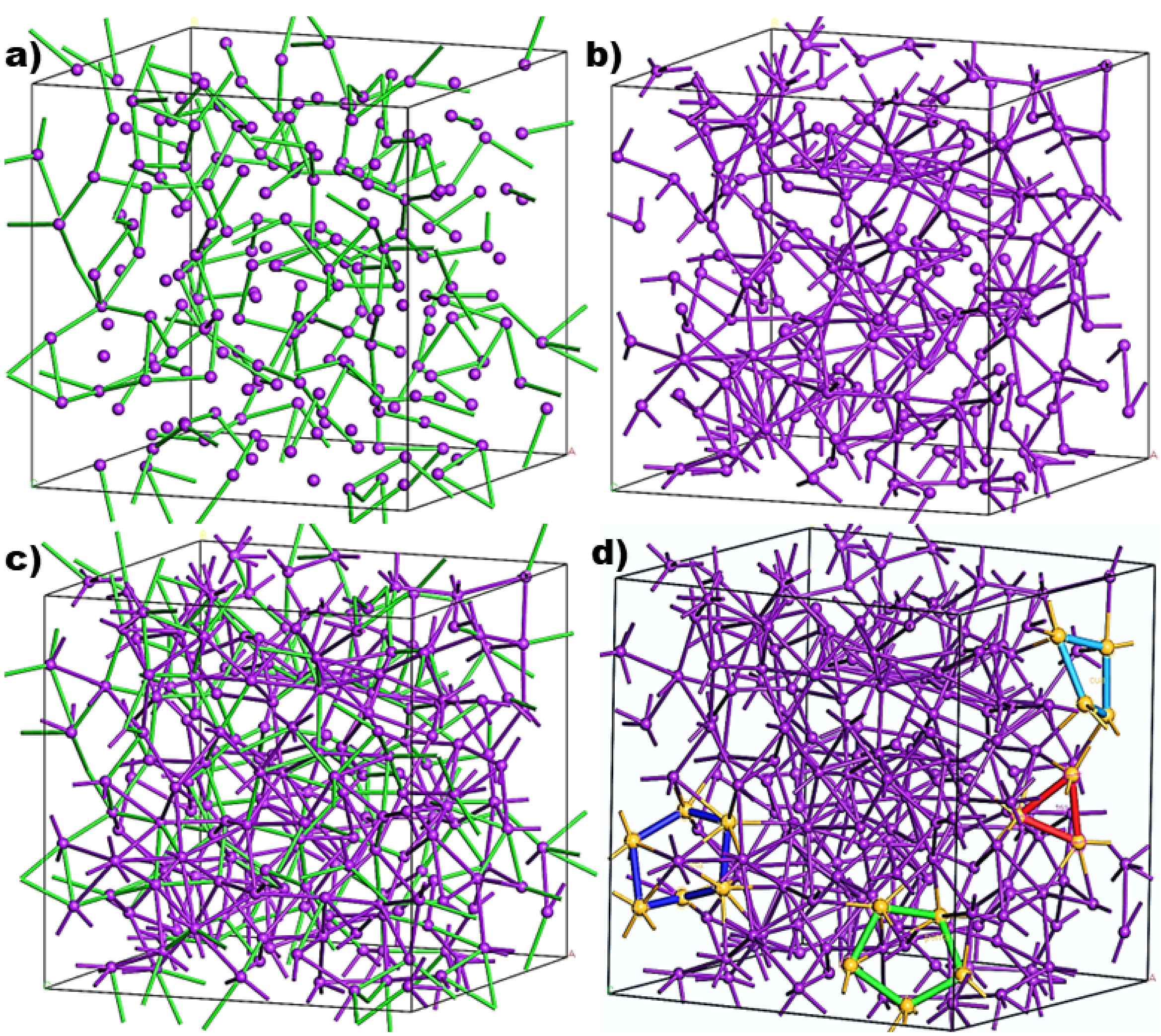

*Figure 11. Atomic structures as a function of the region in the PDFs. a) Region of increasing PDFs in the first peak; b) decreasing region, c) region of the first peak, d) Some atomic structures that appear after completion of the first peak.*

### *3.3 The atomic arrangements in the supercell*

Fig. 11 depicts the evolution of the interatomic bonding as a function of the distance between pairs of atoms. The reference distance is less than or equal to the minimum between the first peak and the shoulder, namely, 4.185 Å. The regions that are considered are a) the region before the maximum of the first peak; b) the region between the maximum of the first peak and the minimum between the first peak and the shoulder; c) the total region of the first peak and d) some atomic structures that resemble triangles, squares, pentagons and hexagons.

As mentioned, there may be atomic structures more complicated than the ones identified, that would give rise to the "plateau" located between 120º and 140º. This region has not been investigated in this work, but it merits a closer look to see its relevance.

## 4. CONCLUSIONS

As stated before, this work aimed at answering the following questions:

Are the PDFs and PADs obtained in this work comparable with the experimental and simulational ones reported in the literature? What are the most common geometrical structures in the liquid, consistent with the PDFs and the PADs calculated? Is the shoulder related to some specific structure? Is this shoulder the result of some experimental approximations when determining the PDFs from the structure factors?

and we have done so; thereby contributing to the understanding of the structures of liquid Bi and their influence on its physical properties.

The answers according to our work are the following:

Are the PDFs and PADs obtained in this work comparable with the experimental and simulational ones reported in the literature?
----- *Our results for the calculated PDFs indicate the existence of a general behavior consistent throughout the whole molecular dynamics on the plateau between 100 steps and 600 steps. They coincide with some experimental results, but above all, they always manifest the existence of a pseudo-peak (shoulder) regardless of the position in the plateau. The general structure indicates the presence of two main peaks and a shoulder in between, in accord with most of the experimental results. No experiments were found for the determination of the PADs, but compared to the other two simulations, we agree with one and disagree with the other.*

What are the most common geometrical structures in the liquid, consistent with the PDFs and the PADs calculated?
----- *Our results suggest the existence of deformed equilateral triangles and deformed squares, coincident with the PADs and the PDFs results. Other more complex structures should exist congruent with the almost constant region between 120º and 140º.*

Is the shoulder related to some specific structure?
----- *The shoulder exists and it is not a consequence of deficient measurements, unlike what is claimed by some experimentalists. Our computer simulations are statistical and the results are a consequence of some clustering of atoms related to the third and fourth crystalline peaks (second neighbors) that retained some presence. Also, the existence of deformed equilateral triangles and square structures lead to the existence of the shoulder, and these are the origins of the pseudo-peak.*

Is this shoulder the result of some experimental approximations when determining the PDFs from the structure factors?
----- *Definitely not; There are no experimental determinations involved in our results and the shoulder always appears in all the steps of the simulation.*

The prominent angles observed in the PADs are: localized around 60º and somewhat localized around 90º, indicating the presence of deformed triangles and squares, as mentioned above. Moreover, there are wide spread angles around 120º-140º (an almost plateau) that should be studied.

To be more precise, by calculating the PADs, one finds noticeable angles of 60º (53° and 58° for MS and RMC, respectively), corresponding to deformed triangular structures; 90º (between 80º and 100º), corresponding to deformed squares; and angles between 110º and 140º, corresponding to higher order geometrical structures. Correspondingly, the peaks of the PDFs are about 3.25 Å, 4.6 Å (shoulder) and 6.55 Å which closely resembles the nearest-neighbor distance in an equilateral triangle, the position of the double crystalline peak (4.52 and 4.72 Å), resembling the distance subtended by an angle of 90º, and the distance subtending an angle of 140º between two next-nearest-neighbor atoms, reminiscent of a nonagon: 0.9397x2x3.5 = 6.58 Å

To conclude, our results seem to unequivocally identify the geometrical structures obtained from our PAD and our PDF calculations.

## ACKNOWLEDGEMENTS

F.B.Quiroga acknowledges SECIHTI for her graduate fellowship. I.Rodríguez and D.Hinojosa-Romero thank SECIHTI and DGAPA-UNAM (POSDOC) for their respective postdoctoral fellowships. A.A.V., R.M.V., and A.V. thank DGAPA-UNAM for continued financial support to carry out research projects under grants No. IN118223 and IN119226. This research was also supported by SECIHTI under Grant No. CBF-2025-G-886. Computational resources were partially provided by the Supercomputing Center of DGTIC-UNAM through the project LANCAD-UNAM-DGTIC131. M. T. Vázquez and O. Jiménez provided the information requested. A. Pompa helped with the maintenance and support of the computing cluster in IIM-UNAM. Also, Prof. J. Juarez helped with the characterization of the structure of mineral Bi.

## AUTHOR CONTRIBUTIONS

This research was conceived and designed by F.B.Q. and A.A.V., with input from R.M.V., A.V., D.H.-R. and I.R. All simulations were performed by F.B.Q. All authors contributed to the discussion and analysis of the results. F.B.Q. and A.A.V. wrote the initial draft of the manuscript, which was subsequently enriched and approved by all co-authors for publication.

## COMPETING INTERESTS STATEMENT

The authors declare no conflict of interest in this work.

## DATA AVAILABILITY STATEMENT

The datasets used and analyzed during the current study are available from the corresponding author on reasonable request.

## REFERENCES

[1] T. Daeneke *et al.*, “Liquid metals: fundamentals and applications in chemistry,” *Chem. Soc. Rev.*, vol. 47, no. 11, pp. 4073–4111, Jun. 2018, doi: 10.1039/C7CS00043J.

[2] X. Zhang, J. Liu, and Z. Deng, “Bismuth-based liquid metals: advances, applications, and prospects,” *Mater. Horiz.*, 2024, doi: 10.1039/D3MH01722B.

[3] C. Qin *et al.*, “Phase Interface Regulating on Amorphous/Crystalline Bismuth Catalyst for Boosted Electrocatalytic $CO_2$ Reduction to Formate,” *ACS Appl. Mater. Interfaces*, vol. 15, no. 40, pp. 47016–47024, Oct. 2023, doi: 10.1021/acsami.3c10011.

[4] J. Pan *et al.*, “Mechanically Robust Bismuth-Embedded Carbon Microspheres for Ultrafast Charging and Ultrastable Sodium-Ion Batteries,” *J. Am. Chem. Soc.*, vol. 147, no. 4, pp. 3047–3061, Jan. 2025, doi: 10.1021/jacs.4c09824.

[5] L. Robison *et al.*, “A Bismuth Metal–Organic Framework as a Contrast Agent for X-ray Computed Tomography,” *ACS Appl. Bio Mater.*, vol. 2, no. 3, pp. 1197–1203, Mar. 2019, doi: 10.1021/acsabm.8b00778.

[6] V. Plechystyy, I. Shtablavyi, S. Winczewski, K. Rybacki, S. Mudry, and J. Rybicki, “Short-range order structure and free volume distribution in liquid bismuth: X-ray diffraction and computer simulations studies,” *Philosophical Magazine*, vol. 100, no. 17, pp. 2165–2182, Sep. 2020, doi: 10.1080/14786435.2020.1756500.

[7] S. Mudry, I. Shtablavyi, U. Liudkevych, and S. Winczewski, “Structure and thermal expansion of liquid bismuth,” *Materials Science-Poland*, vol. 33, no. 4, pp. 767–773, Dec. 2015, doi: 10.1515/msp-2015-0100.

[8] J. Akola, N. Atodiresei, J. Kalikka, J. Larrucea, and R. O. Jones, “Structure and dynamics in liquid bismuth and Bi *n* clusters: A density functional study,” *J. Chem. Phys.*, vol. 141, no. 19, Nov. 2014, doi: 10.1063/1.4901525.

[9] J. Hafner and W. Jank, “Structural and electronic properties of the liquid polyvalent elements. IV. the pentavalent semimetals and trends across the periodic table,” *Phys. Rev. B*, vol. 45, no. 6, pp. 2739–2749, 1992, doi: 10.1103/PhysRevB.45.2739.

[10] W. Klement, A. Jayaraman, and G. C. Kennedy, “Phase Diagrams of Arsenic, Antimony, and Bismuth at Pressures up to 70 kbars,” *Physical Review*, vol. 131, no. 2, pp. 632–637, Jul. 1963, doi: 10.1103/PhysRev.131.632.

[11] J. H. Chen, H. Iwasaki, and T. Ktkegawa, “Crystal structure of the high pressure phases of bismuth bi iii and bi iii’ by high energy synchrotron x-ray diffraction,” *High Press. Res.*, vol. 15, no. 3, pp. 143–158, Jan. 1996, doi: 10.1080/08957959608240468.

[12] A. A. Valladares, I. Rodríguez, D. Hinojosa-Romero, A. Valladares, and R. M. Valladares, “Possible superconductivity in the Bismuth IV solid phase under pressure,” *Sci. Rep.*, vol. 8, no. 1, pp. 1–7, 2018, doi: 10.1038/s41598-018-24150-3.

[13] W. Buckel and R. Hilsch, “Supraleitung und elektrischer Widerstand neuartiger Zinn-Wismut-Legierungen,” *Zeitschrift f� Physik*, vol. 146, no. 1, pp. 27–38, Feb. 1956, doi: 10.1007/BF01326000.

[14] W. Buckel and R. Hilsch, “Einflu�der Kondensation bei tiefen Temperaturen auf den elektrischen Widerstand und die Supraleitung f� verschiedene Metalle,” *Zeitschrift f� Physik*, vol. 138, no. 2, pp. 109–120, Apr. 1954, doi: 10.1007/BF01337903.

[15] O. Prakash, A. Kumar, A. Thamizhavel, and S. Ramakrishnan, “Evidence for bulk superconductivity in pure bismuth single crystals at ambient pressure,” *Science (1979).*, vol. 355, no. 6320, pp. 52–55, 2017, doi: 10.1126/science.aaf8227.

[16] D. Hinojosa-Romero, I. Rodriguez, A. Valladares, R. M. Valladares, and A. A. Valladares, “Ab initio Study of the Amorphous Cu-Bi System,” *MRS Adv.*, vol. 4, no. 2, pp. 81–86, Jan. 2019, doi: 10.1557/adv.2019.83.

[17] L. Powell *et al.*, “Multiphase superconductivity in PdBi2,” *Nat. Commun.*, vol. 16, no. 1, p. 291, Jan. 2025, doi: 10.1038/s41467-024-54867-x.

[18] N. K. Karn *et al.*, “Type-II superconductivity at 9K in Pb–Bi alloy,” *Solid State Commun.*, vol. 391, p. 115639, Nov. 2024, doi: 10.1016/j.ssc.2024.115639.

[19] Y. Mizuguchi *et al.*, "Superconductivity in Novel $BiS_2$ -Based Layered Superconductor $LaO_{1-x}F_xBiS_2$," *J. Physical Soc. Japan*, vol. 81, no. 11, p. 114725, Nov. 2012, doi: 10.1143/JPSJ.81.114725.

[20] K. Górnicka *et al.*, "Superconductivity on a Bi Square Net in LiBi," *Chemistry of Materials*, vol. 32, no. 7, pp. 3150–3159, Apr. 2020, doi: 10.1021/acs.chemmater.0c00179.

[21] P.-F. Liu *et al.*, "Prediction of superconductivity and topological aspects in single-layer βBiPd," *Phys. Rev. B*, vol. 102, no. 15, p. 155406, Oct. 2020, doi: 10.1103/PhysRevB.102.155406.

[22] Z. Wang *et al.*, "Correlating the charge-transfer gap to the maximum transition temperature in Bi 2 Sr 2 Ca n-1 Cu n O 2n+4+d." [Online]. Available: https://www.science.org

[23] X. Cheng, E. E. Gordon, M. Whangbo, and S. Deng, "Superconductivity Induced by Oxygen Doping in $Y_2O_2Bi$," *Angewandte Chemie*, vol. 129, no. 34, pp. 10257–10260, Aug. 2017, doi: 10.1002/ange.201701427.

[24] J. Barzola-Quiquia, C. Lauinger, M. Zoraghi, M. Stiller, S. Sharma, and P. Häussler, "Superconductivity in the amorphous phase of topological insulator $Bi_xSb_{100-x}$ alloys," *Supercond. Sci. Technol.*, vol. 30, no. 1, p. 015013, Jan. 2017, doi: 10.1088/0953-2048/30/1/015013.

[25] J. Park *et al.*, "Superconducting phase diagram in BixNi1-x thin films: The effects of Bi stoichiometry on superconductivity," *Phys. Rev. Mater.*, vol. 8, no. 7, p. 074805, Jul. 2024, doi: 10.1103/PhysRevMaterials.8.074805.

[26] M. V. Likholetova, E. V. Charnaya, E. V. Shevchenko, and Yu. A. Kumzerov, "Superconductivity of the Bi–Sn Eutectic Alloy," *Physics of the Solid State*, vol. 63, no. 2, pp. 232–236, Feb. 2021, doi: 10.1134/S1063783421020153.

[27] Z. Mata-Pinzón, A. A. Valladares, R. M. Valladares, and A. Valladares, "Superconductivity in Bismuth. A New Look at an Old Problem," *PLoS One*, vol. 11, no. 1, p. e0147645, Jan. 2016, doi: 10.1371/journal.pone.0147645.

[28] I. Rodríguez, D. Hinojosa-Romero, A. Valladares, R. M. Valladares, and A. A. Valladares, "A facile approach to calculating superconducting transition temperatures in the bismuth solid phases," *Sci. Rep.*, vol. 9, no. 1, pp. 1–14, 2019, doi: 10.1038/s41598-019-41401-z.

[29] B. N. Galimzyanov, A. A. Tsygankov, A. V. Suslov, V. I. Lad'yanov, and A. V. Mokshin, "Quasi-Stable Structures in Equilibrium Dense Bismuth Melt: Experimental and First Principles Theoretical Studies," *Scr. Mater.*, vol. 235, no. June, p. 115618, Jun. 2023, doi: 10.1016/j.scriptamat.2023.115618.

[30] M. Mayo, E. Yahel, Y. Greenberg, E. N. Caspi, B. Beuneu, and G. Makov, "Determination of the structure of liquids: an asymptotic approach," *J. Appl. Crystallogr.*, vol. 46, no. 6, pp. 1582–1591, Dec. 2013, doi: 10.1107/S002188981302431X.

[31] U. Dahlborg and M. Davidovi, "On the Structure of Liquid Bismuth," *Phys. Chem. Liquids*, vol. 15, no. 4, pp. 243–252, Mar. 1986, doi: 10.1080/00319108608078486.

[32] E. N. Caspi, Y. Greenberg, E. Yahel, B. Beuneu, and G. Makov, "What is the structure of liquid Bismuth?," *J. Phys. Conf. Ser.*, vol. 340, 2012, doi: 10.1088/1742-6596/340/1/012079.

[33] B. Delley, "DMol, a standard tool for density functional calculations: Review and advances," *Theoretical and Computational Chemistry*, vol. 2, no. C, pp. 221–254, Jan. 1995, doi: 10.1016/S1380-7323(05)80037-8.

[34] Dassault Systèmes BIOVIA, "BIOVIA Materials Studio, Release," 2015, *Dassault Systèmes, San Diego*: 2020.

[35] W. G. Hoover, "Canonical dynamics: Equilibrium phase-space distributions," *Phys. Rev. A (Coll. Park).*, vol. 31, no. 3, pp. 1695–1697, Mar. 1985, doi: 10.1103/PhysRevA.31.1695.

[36] S. H. Vosko, L. Wilk, and M. Nusair, "Accurate spin-dependent electron liquid correlation energies for local spin density calculations: a critical analysis," *Can. J. Phys.*, vol. 58, no. 8, pp. 1200–1211, Aug. 1980, doi: 10.1139/p80-159.

[37] I. Rodríguez-Aguirre, D. Hinojosa-Romero, A. Valladares, R. M. Valladares, and Valladares Ariel A., "VallaClan: Computational MAterials Sience Code and Analysis Repository." Accessed: May 19, 2026. [Online]. Available: https://github.com/Isurwars/Vallaclan

[38] M. G. Tucker, D. A. Keen, M. T. Dove, A. L. Goodwin, and Q. Hui, "RMCProfile: reverse Monte Carlo for polycrystalline materials," *Journal of Physics: Condensed Matter*, vol. 19, no. 33, p. 335218, Aug. 2007, doi: 10.1088/0953-8984/19/33/335218.

[39] Y. Zhang, M. Eremenko, V. Krayzman, M. G. Tucker, and I. Levin, "New capabilities for enhancement of *RMCProfile* : instrumental profiles with arbitrary peak shapes for structural refinements using the reverse Monte Carlo method," *J. Appl. Crystallogr.*, vol. 53, no. 6, pp. 1509–1518, Dec. 2020, doi: 10.1107/S1600576720013254.

[40] I. Rodríguez, R. Valladares, A. Valladares, D. Hinojosa-Romero, U. Santiago, and A. Valladares, "Correlation: An Analysis Tool for Liquids and for Amorphous Solids," *J. Open Source Softw.*, vol. 6, no. 65, p. 2976, Sep. 2021, doi: 10.21105/joss.02976.

[41] Y. Waseda and K. Suzuki, "Structure factor and atomic distribution in liquid metals by X-ray diffraction," *physica status solidi (b)*, vol. 49, no. 1, pp. 339–347, Jan. 1972, doi: 10.1002/pssb.2220490132.

[42] Y. Greenberg *et al.*, "Evidence for a temperature-driven structural transformation in liquid bismuth," *Epl*, vol. 86, no. 3, 2009, doi: 10.1209/0295-5075/86/36004.